\documentstyle[emulateapj]{article}
\slugcomment{Accepted AJ, 2005 Sep 21}

\lefthead{MIPS-Ge Observations of the xFLS}
\righthead{Frayer et al.}

\newcommand{\kps}{\,\rm{km~s}^{-1}}
\newcommand{\lsun}{\,L_{\odot}}

\begin{document}

\title{{\it Spitzer} 70$\mu$\lowercase{m} and 160$\mu$\lowercase{m}
Observations of the Extragalactic First Look Survey}

\author{D.\ T.\ Frayer\altaffilmark{1},
D.\ Fadda\altaffilmark{1},
L.\ Yan\altaffilmark{1},
F.\ R.\ Marleau\altaffilmark{1},
P.\ I.\ Choi\altaffilmark{1},
G.\ Helou\altaffilmark{1},
B.\ T.\ Soifer\altaffilmark{1},
P.\ N.\ Appleton\altaffilmark{1},
L.\ Armus\altaffilmark{1},
R.\ Beck\altaffilmark{1},
H.\ Dole\altaffilmark{2},
C.\ W.\ Engelbracht\altaffilmark{3},
F.\ Fang\altaffilmark{1},
K.\ D.\ Gordon\altaffilmark{3},
I. Heinrichsen\altaffilmark{1},
D.\ Henderson\altaffilmark{1},
T.\ Hesselroth\altaffilmark{1},
M.\ Im\altaffilmark{4},
D.\ M.\ Kelly\altaffilmark{3},
M.\ Lacy\altaffilmark{1},
S.\ Laine\altaffilmark{1},
W.\ B.\ Latter\altaffilmark{1},
W.\ Mahoney\altaffilmark{1},
D.\ Makovoz\altaffilmark{1},
F.\ J.\ Masci\altaffilmark{1},
J.\ E.\ Morrison\altaffilmark{3},
M.\ Moshir\altaffilmark{1},
A.\ Noriega-Crespo\altaffilmark{1},
D.\ L.\ Padgett\altaffilmark{1},
M.\ Pesenson\altaffilmark{1},
D.\ L.\ Shupe\altaffilmark{1},
G.\ K.\ Squires\altaffilmark{1},
L.\ J.\ Storrie-Lombardi\altaffilmark{1},
J.\ A.\ Surace\altaffilmark{1},
H.\ I.\ Teplitz\altaffilmark{1},
G.\ Wilson\altaffilmark{1}
}

\altaffiltext{1}{{\it Spitzer} Science Center, California Institute of
Technology 220--06, Pasadena, CA 91125; frayer@ipac.caltech.edu.}  

\altaffiltext{2}{Institut d'
Astrophysique Spatiale, b\^{a}t 121, Universit\'{e} Paris Sud, F-91405
Orsay Cedex, France.}  

\altaffiltext{3}{Steward Observatory,
University of Arizona, 933 North Cherry Avenue, Tucson, AZ 85721.}

\altaffiltext{4}{Astronomy Program, Seoul National University,
Shillim-dong, Kwanak-gu, Seoul, S. Korea, 2-880-9010.}

\begin{abstract}

We present {\it Spitzer} 70$\mu$m and 160$\mu$m observations of the
{\it Spitzer} extragalactic First Look Survey (xFLS).  The data
reduction techniques and the methods for producing co-added mosaics
and source catalogs are discussed.  Currently, 26\% of the 70$\mu$m
sample and 49\% of the 160$\mu$m-selected sources have redshifts.  The
majority of sources with redshifts are star-forming galaxies at
$z<0.5$, while about 5\% have infrared colors consistent with AGN.
The observed infrared colors agree with the spectral energy
distribution (SEDs) of local galaxies previously determined from {\it
IRAS} and {\it ISO} data.  The average 160$\mu$m/70$\mu$m color
temperature for the dust is $T_{d}\simeq 30\pm5$\,K, and the average
70$\mu$m/24$\mu$m spectral index is $\alpha\simeq 2.4\pm0.4$.  The
observed infrared to radio correlation varies with redshift as
expected out to $z\sim 1$ based on the SEDs of local galaxies.  The
xFLS number counts at 70$\mu$m and 160$\mu$m are consistent within
uncertainties with the models of galaxy evolution, but there are
indications that the current models may require slight modifications.
Deeper 70$\mu$m observations are needed to constrain the models, and
redshifts for the faint sources are required to measure the evolution
of the infrared luminosity function.

\end{abstract}


\section{Introduction}

One of the first science observations carried out with the {\it
Spitzer Space Telescope} (Werner et al. 2004) was the non-proprietary
extragalactic First Look Survey (xFLS) which was designed to
characterize the infrared sky at previously unexplored sensitivities.
The {\it IRAS} mission first uncovered the presence of infrared
luminous galaxies in the local universe (Neugebauer et al. 1984;
Soifer et al. 1987), and the {\it ISO} infrared (Elbaz et al. 1999,
2002; Rodighiero et al. 2005) and ground-based submillimeter and
millimeter observations (Blain et al. 2002 and references therein)
have highlighted the importance that infrared luminous galaxies have
on the general understanding of galaxy evolution.  Since the cosmic
infrared background (CIB) peaks in the far-infrared (FIR) (Hauser \&
Dwek 2001), studying the properties of galaxies that are bright in the
FIR is crucial for constraining models of galaxy evolution.

In this paper, we present 70$\mu$m and 160$\mu$m observations of the
xFLS field using the Multiband Imaging Photometer for {\it Spitzer}
(MIPS, Rieke et al. 2004).  The xFLS is a 4\,deg$^2$ survey.  The
SWIRE {\it Spitzer} survey (Londsdale et al. 2004) covers a wider area
(49\,deg$^2$) to similar depths, and deeper observations covering
smaller areas are being taken by the MIPS Instrument Team as part of
the Guaranteed Time Observers (GTO) program (e.g., Dole et al. 2004a)
and other groups.  Although the xFLS is not unique in terms of depth
or area coverage, the field has a large number ($\sim 3000$) of
spectroscopic redshifts (P. Choi et al. 2005, in preparation;
F. Marleau et al. 2005, in preparation), ancillary radio (Condon et
al. 2003), and optical imaging data (Fadda et al. 2004) that permit
detailed multi-wavelength studies over a relatively large area.  In
this paper, we measure the source counts and use the available
redshifts to constrain the rest-frame spectral energy distributions
(SEDs) and derive the average infrared properties for the xFLS
70$\mu$m and 160$\mu$m populations.  A cosmology of H$_0=70\kps\,{\rm
Mpc}^{-1}$, $\Omega_{\rm M}=0.3$, and $\Omega_{\Lambda}=0.7$ is
assumed throughout this paper.

\section{Observations}

The xFLS survey covers a 4\,deg$^2$ region ($17^{\rm h}18^{\rm
m}00^{\rm s}$, $+59^{\circ}30^{\prime}00^{\prime\prime}$) within the
northern continuous viewing zone of {\it Spitzer}\footnote{The
extragalactic FLS data can be retrieved from the {\it Spitzer} Science
Center at http://ssc.spitzer.caltech.edu/fls/}.  Inside the xFLS main
field a smaller verification strip of 0.25\,deg$^{2}$ centered at
$17^{\rm h}17^{\rm m}00^{\rm s}$,
$+59^{\circ}45^{\prime}00^{\prime\prime}$ was observed with an
integration time of 4 times that of the main survey to characterize
the completeness and source reliability of the main survey.  In total,
27.7 hours of xFLS observations were taken in 2003 December with the
MIPS instrument.  An additional 16.8 hours of observations within the
xFLS main field were taken in 2005 May to characterize the performance
of the 70$\mu$m array at warmer telescope temperatures
($T_{mirror}\simeq 9.5$\,K).  These observations produced useful data
at 70$\mu$m, but the 160$\mu$m data were not usable.  Figure~1 shows
the layout of the observations.

All of the MIPS observations were taken using the medium scan-rate
mapping mode (4.2\,s data collection events [DCEs]).  The main-survey
data were taken with adjacent 2\,deg scan legs that were offset in the
cross-scan direction by $276^{\prime\prime}$ (nearly the full field of
view of the arrays).  The main-survey was covered twice to identify
potential asteroids and to increase the redundancy of the data set.
Unfortunately, half of the MIPS-70 array was rendered useless after
launch due to a cable failure outside the instrument (Rieke et
al. 2004).  Therefore, at 70$\mu$m each position of the main survey
was covered by only one scan leg.  Since the data were not scheduled
exactly as originally planned, slight rotations between the 8
astronomical observation requests (AORs) yielded small gaps with zero
coverage in the south-west corner of the main-survey 70$\mu$m map.

The verification strip was observed using 4 AORs with 0.5\,deg scan
legs at the medium scan rate.  Each AOR covered the
0.5\,deg$\times$0.5\,deg field using cross-scan steps of
$148^{\prime\prime}$ (slightly less than half an array field of view).
The warm test data taken in 2005 May were centered on the 70$\mu$m
main-field.  These data consisted of 6 AORs with 1.75\,deg scan legs.
Cross-scan steps of $148^{\prime\prime}$ were used to map the
1.75\,deg$\times$1.75\,deg field once.  Table~1 shows the average
integration times, sensitivities, and area coverage for the 70$\mu$m
and 160$\mu$m observations.

\section{Data Reduction}

\subsection{BCD Pipeline Processing}

The raw 70$\mu$m and 160$\mu$m (MIPS-Germanium [Ge]) data were
downloaded from the {\it Spitzer} Science Center (SSC) archive and
were reduced using the offline Ge Reprocessing Tools (GeRT), available
from the public SSC website.  The GeRT uses an offline version of the
SSC pipeline to produce the basic calibrated data products (BCDs),
following the algorithms derived by the MIPS Instrument Team and the
MIPS Instrument Support Team (Gordon et al. 2005).  The processing was
done using the latest software (SSC pipeline version S12) to take
advantage of improvements not currently available for the online xFLS
data products (made with pipeline version S11).

BCD processing has two main steps: (1) calculation of the slope of the
data ramp, and (2) calibration of the slope image.  For the xFLS
MIPS-Ge data, the raw 4.2\,s data ramps are comprised of 32
non-destructive reads per pixel (one DCE).  After correcting for the
electronic nonlinearity, cosmic ray events and other discontinuities
are identified in the data ramps.  The SSC pipeline identifies
discontinuities using a maximum likelihood technique (Hesselroth et
al. 2000).  Linear slopes are calculated for the ramp segments between
discontinuities and checked for consistency.  The slopes from the
segments are combined based on the empirical errors estimated from the
scatter of the data within the ramp segments.  Average slopes for each
pixel are calculated to produce an uncalibrated slope image.

The second step of BCD processing is the calibration of the slope
image.  The calibration of MIPS-Ge data is based on frequent
measurements of internal stimulator flashes (stims) which are used to
track the responsivity of the detectors as a function of time.  For
the xFLS medium scan-rate data, the stim flashes are observed every 25
DCEs (118\,s).  The stim flash signal is measured by subtracting the
previous ``background'' DCE from the stim frame which is taken at the
same position on the sky.  For each AOR, a stim response function
($SR[t]$) is calculated from interpolating between the stim minus
background measurements.

After the determination of the stim response as a function of time
 ($SR[t]$), the BCD data are calibrated using the following equation:
\begin{equation}
BCD(t)=FC[U(t)/SR(t) - DARK]/IC,
\end{equation}
where U(t) is the uncalibrated slope image, DARK is the dark
calibration file, and IC is the illumination correction calibration
file which corrects for the combined illumination pattern from the
telescope and the stim flash signal.  The DARK and IC calibration
files are stable and are generated by combining data from several
different campaigns to improve the signal-to-noise.  The flux
conversion factor ($FC$) converts the instrument units into physical
surface brightness units of MJy\,sr$^{-1}$ and is derived from
observations of standard calibrators (Sec. 3.4).

\subsection{BCD Filtering}

Before the 70$\mu$m bias level was lowered in 2004 March, the 70$\mu$m
data showed significant data artifacts.  Examples and discussion of
MIPS-Ge artifacts are shown in the MIPS Data
Handbook\footnote{http://ssc.spitzer.caltech.edu/mips/dh/}. The two
main artifacts impacting the xFLS data are the stim flash latents and
the variations of the slow response as a function of time.  The
160$\mu$m data are affected by these issues to a lesser degree due to
the faster time constants of the 160$\mu$m stressed-Germanium
detectors.  For point sources, the stim latent and slow response
residuals are additive effects.  The filtering of the 160$\mu$m data
is straight-forward.  We used the filtered BCDs products, which remove
a running median per pixel as a function of time by subtracting the
median value for the surrounding DCEs closest in time (ignoring the
current DCE, stim DCEs, and bad data).  This high-pass time median
filter removes data artifacts as well as the extended cirrus emission.
To remove the stim latent artifacts at 160$\mu$m, we simply ignored
the first DCE after the stim flash since the stim latents decay away
within one DCE.

The filtering process for the 70$\mu$m data is slightly more
complicated.  The time median filter by itself does not remove all of
the data artifacts.  Stim latents remain for many DCEs and are
correlated by column.  Since the scan map direction is nearly along
the columns of the array (along the y-axis), the column artifacts are
amplified.  We remove the column residuals by subtracting the median
of the values along each column for every BCD.  The combination of the
high-pass time median filter per pixel and the column median filter
removes the bulk of the data artifacts at 70$\mu$m.

The resultant rms is lower for narrow high-pass median filter widths
($<15$ DCEs), but narrow filter windows can yield significant negative
side-lobes around bright sources.  In addition, column filtering does
not maintain point source calibration for the brightest sources ($\ga
0.5$\,Jy).  To avoid significant negative side-lobes and to preserve
calibration, the filtering of the MIPS-Ge data was done in two passes.
The data from the first filtering pass were co-added, and sources were
extracted (Sec. 3.3) to find the location of the bright sources.  The
source positions within the original BCDs were masked and new
filtering corrections were calculated in the second pass, ignoring the
pixels containing sources.  This second-pass filtering technique
minimizes the data artifacts while preserving point-source
calibration.

Different filtering methods were tested and optimized using the GeRT.
At 70$\mu$m, we applied column filtering followed by a high-pass time
median filter with a width of 12 DCEs to yield the deepest image.  The
final sensitivity of the filtered 70$\mu$m image was improved by more
than a factor of two in comparison to data with no filtering.
Filtering has less of an impact at 160$\mu$m, but is useful for deep
observations of faint point sources.  We adopted a high-pass time
median filter width of 12 DCEs for the 160$\mu$m data using the
second-pass filtering technique as done for 70$\mu$m.

\subsection{Data Coaddition and Source Extraction}

The filtered BCDs were coadded using the SSC mosaicing and source
extraction software (MOPEX) (Makovoz \& Marleau 2005).  The data were
combined ignoring bad data flagged in the BCD mask files (bmask) and
the bad pixels defined by the static pixel mask (pmask).  Using the
redundancy of the MIPS data, additional spurious data values were
removed via outlier rejection.  The remaining data were corrected for
array distortions and projected onto a sky grid with square pixels
that were more than 2 times smaller than the original pixels.  The
output mosaic pixels are $4\arcsec$ at 70$\mu$m and $8\arcsec$ at
160$\mu$m.  The data were averaged using weights proportional to the
fractional area subtended by the original pixels as projected onto the
output pixel grid.  Since the calibration of the main, verification,
and the warm 70$\mu$m data sets are consistent within measurement
errors, all data were combined to yield the deepest images.

Sources were extracted from the final images using MOPEX.  The MOPEX
software uses a Point-source Response Function (PRF) image to optimize
source detection and point source fitting.  A PRF calibration image
was made for each MIPS band by coadding isolated bright point sources
within the xFLS field.  A point source probability (PSP) image was made
from non-linear matched filtering of the input image with the PRF.
The PSP image represents the probability of having a point source
above the noise at each pixel.  The PSP image was multiplied by the
input image to yield the filtered image used for source detection.
The filtered image enhances the presence of point sources in the
mosaic while smoothing out noise features that do not match the input
PRF, providing a more robust image for source detection.  After source
detection, the original images were fitted with the PRF image via
$\chi^2$ minimization to extract source positions and flux densities.
Regions containing multiple peaks were de-blended.  For extended
sources and/or close blends, not well fitted by the PRF, large
aperture measurements were used to derive the total flux density for
the source.

The noise across the images is non-uniform.  For optimal source
detection and extraction, an accurate representation of the
uncertainties is needed.  The MOPEX uncertainty image based on the
input BCD pipeline uncertainties is useful for identifying the
relative uncertainties across the image, but the pipeline errors
typically underestimate the absolute uncertainty level for low
background regions.  The median level of the MOPEX uncertainty image
was scaled to match the average empirical noise in the mosaics.  By
using a properly scaled uncertainty image, we find an average $\chi^2$
for the fitted point sources of $\chi^2\simeq 1$, suggesting that the
uncertainty image is self-consistent with the average point source
errors in the data.

Several tests were done to validate the photometric results from
MOPEX.  No calibration differences were found for different sub-sets
of the data.  We also compared the photometry done with the MOPEX
software with that from the program StarFinder (Diolaiti et al. 2000).
Both programs use an empirical PRF to fit source flux density.  The
results from MOPEX and StarFinder agree to better than 5\% for bright
point sources ($\ga 50$\,mJy at 70$\mu$m and $\ga$\,200\,mJy at
160$\mu$m) and better than 10\% rms down to the limits of the catalog,
consistent with the uncertainties in the data.  For faint sources
detected in the MIPS-Ge bands, PRF fitting techniques yield more
reliable results than aperture measurements.  By comparing results
between the verification data and the main survey, we find a
consistency of $\pm10$\% for PRF matching techniques and a dispersion
for aperture measurements of larger than $\pm25$\% at faint flux
density levels.

\subsection{Calibration}

The absolute calibration is derived using observations of stars at
70$\mu$m and observations of asteroids and well-studied luminous
infrared galaxies at 160$\mu$m.  The data were processed assuming flux
conversion factors of 634 MJy\,sr$^{-1}$ per MIPS-70 data unit and 42
MJy\,sr$^{-1}$ per MIPS-160 data unit.  The flux densities in the
source catalogs were multiplied by the correction factors of 1.16 and
1.17 for the MIPS-70 and MIPS-160 bands respectively, which includes
the color corrections associated with a constant $\nu f_{\nu}$ SED and
updated calibration conversion factors based on the latest
measurements.  Additional color corrections for sources with
non-constant $\nu f_{\nu}$ SEDs are negligible since the color
corrections are similar for the range of SEDs appropriate for
galaxies; $f_{\nu}\propto \nu^{-\alpha}$, where $\alpha \sim
0$\,--\,3.  For flux density ranges of 50\,mJy\,--\,2\,Jy, the
absolute calibration uncertainty is estimated to be about 15\% and
25\% for the 70$\mu$m and 160$\mu$m bands respectively.  No
corrections have been made for possible flux nonlinearities, which may
be significant for sources brighter than this nominal flux density
range.  Observers are recommended to check the latest information on
MIPS calibration at the SSC web site.

We confirmed that the absolute 70$\mu$m flux density scale for the
xFLS data is consistent with previous measurements made by {\it IRAS}
to within 10\%.  For comparison with {\it IRAS}, we interpolated
between the flux densities in {\it IRAS}-60 and {\it IRAS}-100 bands
(Beichman et al. 1988) to the effective wavelength of the MIPS-70 band
(71.4$\mu$m)\footnote{The effective wavelengths of the MIPS-24, 70,
and 160 bands are 23.7$\mu$m, 71.4$\mu$m, and 155.9$\mu$m,
respectively.  Throughout this paper, the flux densities for the
MIPS-bands are defined as S24$\equiv$ S$_\nu$(23.7$\mu$m), S70$\equiv$
S$_\nu$(71.4$\mu$m), and S160$\equiv$ S$_\nu$(155.9$\mu$m).} applying
the appropriate color corrections.  For the five brightest sources in
the field detected by {\it IRAS}, we derive a flux density ratio of
{\it Spitzer/IRAS} $=1.0\pm0.1$ at 71.4$\mu$m.  The calibration at
160$\mu$m is more uncertain given that xFLS sources have not been
observed previously at wavelengths longer than 100$\mu$m.  Simple SED
model fits to the data indicate that the calibration at 160$\mu$m is
consistent with {\it IRAS} to within 30\%.

\subsection{Mosaics and Catalogs}

The MIPS-Ge mosaics and catalogs are available online at the SSC
website.  For both of the MIPS-70 and 160$\mu$m bands, we provide
coadded mosaics of the entire xFLS data sets along with the
corresponding coverage maps and uncertainty images.  The mosaics are
from the filtered products and have been background subtracted.  The
absolute level of the uncertainty image has been scaled to match the
empirical noise within the mosaics.  The coverage map represents the
effective number of input BCDs for each pixel of the mosaic after
outlier rejection.  There are regions within the 70$\mu$m and
160$\mu$m mosaics with zero coverage.  The 70$\mu$m image has gaps in
the south-west corner of the mosaic due to non-overlapping AORs.  The
160$\mu$m mosaics have a regular pattern of low coverage due to the
unusable block of five detectors and the low redundancy of these
observations.

Point-source catalogs were made for sources with S/N$>7$ at each
wavelength to insure high reliability.  Sources near the edges or
within regions of very low coverage were deleted from the public
catalogs to avoid potentially spurious sources.  In total, 687
70$\mu$m sources (Table~2) and 207 160$\mu$m sources (Table~3) are
cataloged.  The 70$\mu$m and 160$\mu$m catalogs are independent.  The
catalogs can be combined together or with xFLS data at other
wavelengths, depending on the specific application.  The average point
source rms ($1\sigma$) is 2.8\,mJy and 20\,mJy for the 70$\mu$m
warm+main data and 160$\mu$m main-survey data, respectively.  For the
deeper verification regions, the point source rms is 1.6\,mJy at
70$\mu$m and 10\,mJy at 160$\mu$m.

The images were examined visually to validate source detection and to
determine the appropriate method for deriving the flux densities on a
source by source basis.  About 98\% of the derived flux densities are
based on PRF fitting (Sec.\,3.3).  The remaining sources mostly
consist of bright extended galaxies and/or close blends that are not
well fitted by the PRF.  In these cases, large aperture measurements
were made to derive the total flux densities.  For bright extended
sources that have one or more neighboring point sources within the
aperture, the flux density of the bright galaxy was derived by
subtracting the PRF measurement(s) of the faint point source(s) from
the total aperture measurement.

Table~2\&3 show the format for an example portion of the xFLS MIPS
70$\mu$m and 160$\mu$m catalogs published in the online edition of the
Journal.  The average radial positional errors are $2\farcs6$ and
$5\farcs2$ ($1\sigma$) for the S/N$>7$ MIPS-70 and 160$\mu$m sources
respectively.  No systematic differences are found for the 70$\mu$m
source coordinates ($<0\farcs2$) in comparison to the more accurate
24$\mu$m positions.  For consistency with the 24$\mu$m and 70$\mu$m
data, we correct the 160$\mu$m positions by $4\farcs7$ to compensate
for the systematic positional offset measured for the 160$\mu$m
sources (a known issue for pre-S12 versions of the pointing pipeline).
The flux densities have been color corrected assuming a constant $\nu
f_{\nu}$ SED.  The tabulated flux density errors include the absolute
flux density uncertainties of 15\% and 25\% at 70$\mu$m and 160$\mu$m,
respectively.

\subsection{Reliability and Completeness}

The verification data were used to test the reliability of the source
detection and extraction techniques.  For a high level of reliability
at $S/N < 10$, it is important that the uncertainty image accurately
reflects the small-scale spatial variations in the noise across the
images.  Using the uncertainties based on the BCDs, properly scaled to
represent the average empirical noise in the data (Sec.\,3.3), we
obtained good results.  Based on the deeper verification data, we do
not find spurious sources with S/N $>5$ in the main survey within the
verification field.  Sources detected at S/N $>4$ are 80-85\%
reliable, and detections at the S/N $>3$ level are only 50-60\%
reliable.  Based on these results, we adopted a S/N $>5$ cut for
deriving the source counts in Sec.\,4.1.  A more conservative S/N$>7$
criterion was adopted for the public catalogs (Sec. 3.5).

At 70$\mu$m the completeness and source counts were measured within
the overlapping region of the warm and main fields (Fig. 1), while the
entire main-field xFLS area was used at 160$\mu$m.  Both empirical
completeness estimates based on the verification data and simulated
completeness measurements were made.  The simulations of completeness
were carried out by adding point sources with different flux densities
into the mosaics at random locations, and then extracting sources
using the same techniques adopted for the source catalogs.  The
simulations and empirical methods for estimating completeness gave
consistent results.  Using the adopted S/N$>5$ criterion, the
completeness level falls rapidly below 60\% for S70$<14$\,mJy (Fig. 2)
and S160$<100$\,mJy (Fig. 3).  Similar completeness simulations were
done for the deeper verification field, and we find 60\% completeness
levels of approximately 9\,mJy and 60\,mJy at 70$\mu$m and 160$\mu$m,
respectively.

\section{Results and Discussion}

\subsection{Source Counts}

The source counts are derived for the main xFLS and verification
fields based on S/N$>5$ catalogs corrected for completeness.  At
70$\mu$m we use the 3.3 deg$^2$ region containing both the main and
warm survey data.  The 160$\mu$m counts are based on the entire 4.5
deg$^2$ area of the main survey.  Figures 4\&5 show the differential
source counts ($dN/dS\times S^{2.5}$) at 70$\mu$m and 160$\mu$m
respectively.  Over the entire observed range of flux density, both
the 70$\mu$m and 160$\mu$m counts increase at super-Euclidean rates
with decreasing flux density.

In total 845 sources at 70$\mu$m were detected ($>5\sigma$ at
S70$<440$\,mJy) within the main$+$warm field area and 186 sources were
detected (S70$<70$\,mJy) within the verification field (Table~4).  In
comparison at 160$\mu$m, 227 sources were detected in the main-field
(S160$<880$\,mJy) and 45 sources (S160$<140$\,mJy) were found in the
verification field (Table~5).  The source counts are consistent
between the main and verification data for the overlapping range of
flux densities.  The xFLS counts are also consistent within errors
with the previous MIPS-Ge counts published by Dole et al. (2004a).
The error bars for the source counts include the Poisson noise, the
completeness errors, and the absolute flux density calibration
uncertainty.  The data were binned such that the errors are not
dominated by Poisson statistics, except for the highest flux density
bins.  In the lowest flux density bins, the errors on completeness
dominate the uncertainties.

Dole et al. (2004a) found a difference in the 70$\mu$m counts between
the CDF-S and Marano observations, which they attributed to field
variations.  The measured 70$\mu$m xFLS counts are within the range of
values previously measured, and closer to the CDF-S counts at faint
flux densities.  The xFLS-Ge counts are also consistent within
uncertainties with the evolutionary models of Lagache et al. (2004),
which were updated to match the {\it Spitzer} 24$\mu$m counts.
However, the xFLS data points are systematically slightly lower than
the predictions from current models.  Counts from the wide-area SWIRE
survey would be needed to check for possible field variations and to
test whether the models require slight modifications.

Although the xFLS 70$\mu$m counts are at fainter levels than those
previously published (Dole et al. 2004a), they are not yet deep enough
to measure the location of the expected turn over of the differential
counts (Fig. 4).  The flux density at which the differential counts
turn over provides important constraints on evolutionary models
(Lagache et al. 2004; Chary \& Elbaz 2001; King \& Rowan-Robinson
2003; Xu et al. 2001), as shown by the 24$\mu$m data (Marleau et
al. 2004; Papovich et al. 2004; Chary et al. 2004).  At 160$\mu$m {\it
Spitzer} is not expected to measure the turn over in the counts
(Fig. 5), since the 160$\mu$m data are expected to be limited by
confusion below 40\,mJy (Dole et al. 2004b).  The differential counts
at 70$\mu$m are expected to turn over at around 10\,mJy (Lagache et
al. 2004), which is well above the confusion limit, suggesting that
deeper 70$\mu$m counts would provide useful constraints for galaxy
evolution models.

\subsection{Infrared Colors}

For high reliability, the source counts were derived using a S/N$>5$
criterion for the individual bands.  However, it is possible to
maintain a high level of reliability at lower S/N by using the
MIPS-24$\mu$m and/or radio data of the field.  For studying the
infrared colors, we used S/N$>4$ source lists to increase the number
of sources detected at high-redshift in the MIPS-Ge bands.  The
70$\mu$m source list was bandmerged with the 24$\mu$m catalog
(D. Fadda et al., in preparation).  At S/N$>4$, 1779 xFLS-70 sources
have counterparts at 24$\mu$m within $6\arcsec$.  The different
resolutions of the 24$\mu$m, 70$\mu$m, and 160$\mu$m data sets
($6\arcsec$, $18\farcs5$, and $40\arcsec$, respectively) complicate
the identification of the counterparts in the MIPS bands.  Out of the
1779 70$\mu$m sources, 9\% have multiple candidate 24$\mu$m
counterparts.  To avoid potential spurious matches, we only used
sources with one-to-one matches for studying the infrared colors.  The
$4\sigma$ xFLS-160 source list was matched to the 70$\mu$m data set
using a positional matching radius of $13\arcsec$.  The resulting
matched MIPS-160+70 source list was matched to the MIPS-70+24 source
list.  We find 301 160$\mu$m sources with one-to-one matches between
all three MIPS-bands.

Out of the 1618 xFLS-70 sources with one 24$\mu$m positional match,
427 currently have redshifts.  Approximately 57\% of these sources
come from the NOAO WIYN-Hydra radio-selected redshift survey
(F. Marleau et al. in preparation), 16\% of the redshifts are from the
Keck-DEIMOS 24$\mu$m-selected sample (P. Choi et al., in preparation),
and the remaining 27\% of redshifts are from the SLOAN survey (Strauss
et al. 2002).  The largest redshift to date is for a quasar at
$z=3.56$.  For the 160$\mu$m-selected sources with detections in all
three MIPS-bands, about half (146) have known redshifts. Although the
redshift survey of xFLS-Ge sources is not nearly complete (only 26\%
of the 70$\mu$m sources and 49\% of the 160$\mu$m sources currently
have redshifts), there are a sufficient number of sources to study the
general trends in the infrared colors as a function of redshift for
the sample.

We adopt two sets of SED models for studying the infrared colors.  The
first set of SED models is based on a simple modified blackbody
(Blain, Barnard, \& Chapman 2003).  The SED is expressed as $f_{\nu} =
\epsilon_{\nu}B_{\nu}$, where $B_{\nu}(T_{d},\nu)$ is the blackbody
function for a dust temperature $T_{d}$, and $\epsilon_{\nu} \propto
\nu^{\beta}$ is the dust emissivity.  In the mid-IR, we substitute a
power-law of the form $f_{\nu} \propto \nu^{-\alpha}$, smoothly
matching $\epsilon_{\nu}B_{\nu}$ at longer wavelengths (Blain et
al. 2003).  For this simple SED model, there are three free
parameters: $T_{d}$, $\beta$, and $\alpha$.  The S160/S70 ratio
constrains $T_{d}$, while the S70/S24 ratio measures $\alpha$.  The
dust emissivity index $\beta$ is not very well constrained by the {\it
Spitzer} data alone.  For simplicity, we adopt a constant value of
$\beta=1.5$ which is consistent with the results for low-redshift {\it
IRAS} galaxies (Dunne et al. 2000).

The second set of SED models uses a physically more realistic approach
by assuming a power-law distribution of dust masses as a function of
radiation field intensity (Dale et al. 2001).  The dust mass ($M_d$)
as a function of the radiation field ($U$) is given by $dM_{d}(U)
\propto U^{-\gamma} dU$ (Dale et al. 2001; Dale \& Helou
2002)\footnote{We use $\gamma$ to represent the ``$\alpha$'' parameter of
Dale et al. (2001).}.  Quiescent, cirrus-like dust regions are
expected to have $\gamma\simeq 2.5$, while environments near active
H{\sc ii} regions are expected to be approximated by $\gamma\simeq1$.
For a mixture of active and quiescent regions, the average effective
$\gamma$ of star-forming galaxies should be between $1\la \gamma \la
2.5$.

Figure 6 shows the observed 70$\mu$m/24$\mu$m flux density ratio
(S70/S24) plotted as a function of redshift with several predictions
based on the SEDs of local galaxies.  The majority of the xFLS sources
have S70/S24 ratios within the range of values between the extreme
ultraluminous starburst of Arp\,200 and the more typical starburst
M82.  At moderate and high redshifts, infrared-cool ultraluminous
infrared galaxies (ULIRGs, $L>10^{12}\lsun$), e.g., Arp\,220, are
expected to have the highest S70/S24 ratios, while warm-infrared
ULIRGs, e.g., Mrk 231, and AGN dominated sources are expected to have
lower S70/S24 ratios.  Spiral galaxies without a strong infrared
excess in their SED (e.g., M100) show decreasing S70/S24 ratios as a
function of redshift; however, these galaxies are not detected at high
redshifts ($z>0.5$) in the xFLS-Ge survey.  In general, the S70/S24
ratio can be used to help distinguish between AGN-dominated sources
from star-forming galaxies, except in cases for which strong
polycyclic aromatic hydrocarbon (PAH) emission features are redshifted
into the 24$\mu$m band (e.g., M82 at $z\sim2$ in Fig. 6) and for AGN
with strong silicate absorption features (e.g., Spoon et al. 2004) at
$z\sim1.5$.

For the xFLS sample currently with redshifts, only 5\% show low
S70/S24 ratios consistent with AGN SEDs.  The AGN population is fitted
with $T_d=90\pm30$\,K and $\alpha=1.1\pm0.3$ (Fig. 6), which is
consistent with the {\it IRAS} observations of Seyfert galaxies
(Miley, Neugebauer, \& Soifer 1985).  The majority of the xFLS sample
with redshifts are starbursts at $z<0.5$.  Figure 6 shows a large
concentration of sources at $z\sim 0.2$.  The S70/S24 ratios based on
the SEDs of quiescent and active galaxies (Dale \& Helou 2002)
intersect at $z\sim0.2$, which may contribute to this concentration of
data points.  For the galaxies with $z\simeq0.2$, the average observed
infrared color is S70/S24$=14\pm5$.  In comparison with the {\it IRAS}
population (Soifer et al. 1989; Sanders et al. 2003), the {\it
IRAS}-60/{\it IRAS}-25 ratios correspond to a predicted ratio of
S70/S24$=14\pm3$, consistent with the measured xFLS infrared colors.
For the simple SED model, the average S70/S24 ratio corresponds to an
infrared index of $\alpha=2.4$.  Using SEDs of Dale \& Helou (2002),
values of $\gamma \sim 2$ give the best agreement with the average
S70/S24 ratios.

Although the majority of the 70$\mu$m sources do not have redshifts,
the distribution of the S70/S24 ratios is the same for galaxies with
and without redshifts (Fig. 7).  The average mid-IR spectral index is
$\alpha=2.4\pm0.4$.  For a simple power-law representation of the
mid-IR SED, the modeled S70/S24 ratio is constant with redshift (e.g.,
Starburst solid line in Fig. 6).  However, galaxies are expected to
show significant variations in the S70/S24 ratio as strong PAH emission
and mid-IR absorption features (e.g., Armus et al. 2004) are
redshifted through the 24$\mu$m band (as seen for Arp 220 in Fig. 6).

Using simple modified blackbody SEDs, we calculate S160/S70 as a
function of redshift and dust temperature for a constant value of
$\alpha=2.4$ (Fig. 8).  The average 160$\mu$m/70$\mu$m color
temperature for the dust temperature is 30$\pm$5\,K.  This temperature
is slightly lower than the average temperature of $T_{d} = 38\pm3$\,K
(Dunne et al. 2000) and $T_d\simeq 35$--40\,K (Soifer et al. 1989)
derived for {\it IRAS} galaxies, but the longer wavelength 160$\mu$m
band is more sensitive to regions with cooler dust temperatures than
those observed by {\it IRAS}.  The blackbody temperature of 30\,K
derived here agrees well with the long-wavelength {\it ISO}
observations of infrared-bright spiral galaxies (Bendo et al. 2003).
Bendo et al. (2003) found that $T_{d}\simeq 30$\,K provided the best
fit to the {\it ISO} data for $\beta\sim$1--2, consistent with the
results in Figure 8.  In the context of the SED models of Dale \&
Helou (2002), values of $\gamma \sim 2$--2.5 provide the best
agreement with the observed S160/S70 ratios.

The distribution of the S160/S70 ratios is measurably different for
sources with and without redshifts (Fig. 9).  Unlike the S70/S24 ratio
which is on average roughly constant with redshift, the S160/S70 ratio
is expected to increase with redshift (Fig. 8).  Figure~9 shows an
excess of 160$\mu$m sources without redshifts having high S160/S70
ratios (Log[S160/S70]$\ga0.8$).  These results may suggest that the
excess sources are at high redshift ($z\ga 0.5$) and that the current
redshift sample of 160$\mu$m sources may be biased toward lower
redshifts.

\subsection{Infrared Luminosity}

For the 70$\mu$m selected sources, we can estimate the average
infrared luminosity for the xFLS population.  The rest-frame
wavelength of the MIPS-70 band corresponds to 60$\mu$m at a redshift
of $z\simeq0.2$.  The FIR luminosity (42.5--122.5$\mu$m) can be
derived from the rest-frame 60$\mu$m and 100$\mu$m flux densities
(Helou, Soifer, \& Rowan-Robinson 1985).  By using the SED associated
with the average model parameters of $T_d\simeq30$\,K and
$\alpha\simeq2.4$ for the xFLS sources, we estimate a rest-frame
S100/S60 ratio of 2.3.  This is consistent with the values of
S100/S60$=2\pm1$ observed for the bulk of the {\it IRAS} galaxies
(Sanders et al. 2003).  Based on the rest-frame S100/S60 ratio, the
corresponding FIR flux is FIR(W\,m$^{-2})=6.1\times
10^{-14}$\,(S70/Jy) for the observed S70 flux density at $z\simeq0.2$.
For galaxies at $z\simeq0.2$, the average value of S70 is 33\,mJy.
This corresponds to a FIR luminosity of $L$(FIR)$=6.0\times
10^{10}\lsun$.  Using the models of Dale et al. (2001) and the
estimated S100/S60 ratio, the total infrared luminosity
(3--1100$\mu$m) $L$(TIR)$=2.3L$(FIR).  Hence, the average total
infrared luminosity of the 70$\mu$m-selected xFLS galaxies at $z\simeq
0.2$ is about $L$(TIR)$=1.4\times 10^{11}\lsun$.

Although the redshift survey for the 70$\mu$m-selected sources is not
currently complete, the infrared luminosity function can be estimated
at low redshift for bright flux densities.  At S70$>50$\,mJy, 72\%
(65/90) of sources have redshifts.  The majority of sources without
redshifts are suspected to be at high redshift given that the observed
redshift distribution declines significantly at $z>0.3$ and the models
predict that the majority of sources at these flux density levels are
at $z>0.3$.  Based on the models of Lagache et al. (2004), about
40--45\% of galaxies with S70$>50$\,mJy are predicted to be at
$z<0.3$.  Assuming $\sim100$\% completeness at $z<0.3$ (i.e., assume
all sources with S70$>50$\,mJy currently without redshifts are at
$z>0.3$), we find 64$\pm9$\% (58/90) of galaxies with S70$>50$\,mJy
and $z<0.3$.  If the current xFLS redshift surveys are not complete at
$z<0.3$ (with S70$>50$\,mJy), then the models would be even more
discrepant with the observations.  Hence, the models may under predict
the percentage of low redshift 70$\mu$m sources at bright flux
densities.  There are two potential caveats on this result.  First, a
small percentage ($<1$\%) of 70$\mu$m sources do not have 24$\mu$m
counterparts, and these sources are likely to be at high redshift
(e.g., at $z\sim1.5$ where the silicate absorption feature is
redshifted into the 24$\mu$m band).  However, there is only one
70$\mu$m source without a 24$\mu$m counterpart with S70$>50$\,mJy, so
this population, by itself, cannot account for the apparent
discrepancy with the models.  The second caveat is that we have thrown
out 9\% of the 70$\mu$m sources with multiple possible 24$\mu$m
matches (Sec. 4.2).  The 70$\mu$m sources with multiple candidate
24$\mu$m counterparts tend to be brighter on average and typically
have low redshifts.  In fact, we find only one such source with
$z>0.3$ and with S70$>50$\,mJy (1/70 with redshifts).  Additional
redshift surveys are needed to confirm the findings here which suggest
that the current models slightly over predict the percentage of bright
70$\mu$m sources (S70$>50$\,mJy) at high-redshift ($z>0.3$) and their
contribution to the total CIB.

To estimate the infrared luminosity function, we use the sample of 58
galaxies with S70$>50$\,mJy and $z<0.3$.  We adopt the $1/V_{max}$
method following the calculation of the {\it IRAS} $\nu
L_{\nu}(60\mu$m) luminosity function (Soifer et al. 1987).  For each
source, the maximum volume ($V_{max}$) at which the source would be
included in the sample is computed, using $V_{max}(z=0.3)$ as an upper
bound for the survey.  The 70$\mu$m flux density is converted to a
rest-frame 60$\mu$m flux density using a power-law interpolation
between the observed 24$\mu$m and 70$\mu$m measurements.  The
calculated xFLS 60$\mu$m luminosity function ($\rho$) shown in
Figure~10 represents the space density of galaxies per Mpc$^3$ per
magnitude in luminosity.  Even though the current xFLS sample is a
factor of 100 deeper in flux density than the {\it IRAS} 60$\mu$m
bright galaxy sample (Soifer et al. 1987; Sanders et al. 2003), the
xFLS sample does not yet probe the high luminosity end of the
luminosity function (Log($\nu L_{\nu}[60\mu$m$]/\lsun)>11.5$), given
the low redshift range (median redshift of only $z=0.096$) and small
area of the current sample.  The derived xFLS luminosity function
agrees well with the {\it IRAS} 60$\mu$m luminosity function (Soifer
et al. 1987).  Below Log($\nu L_{\nu}[60\mu$m$]/\lsun)<10.5$, $\rho
\propto (\nu\,L_{\nu})^{-0.8}$ (Soifer et al. 1987), while at higher
luminosities $\rho \propto (\nu\,L_{\nu})^{-2.2}$ (Sanders et
al. 2003).  Additional redshifts for the faint S70$\sim$10\,mJy
sources are needed to constrain the evolution of the luminosity
function out to $z\sim1$.

\subsection{Infrared to Radio Correlation}

In the local universe, galaxies over a wide range of luminosity and
Hubble types follow the empirical IR/radio correlation (Helou et
al. 1985; Condon 1992).  Observations of {\it ISO} sources (Gruppioni
et al. 2003) indicate that the mid-IR to radio relationship holds out
to $z\sim 0.6$, and the data for the SCUBA sources suggest that the
FIR to radio relationship may be applicable even at $z\sim 2$--3
(e.g., Chapman et al. 2005).  With {\it Spitzer} we can measure the
IR/radio relationship out to $z\sim 1$ by direct observations near the
peak of the SED, instead of estimating the IR luminosity from the
shorter mid-IR or longer sub-mm wavelengths.  Appleton et al. (2004)
presented the first estimate of the IR/radio relationship based on
{\it Spitzer} data.  Their results were based on the early analysis of
the xFLS data.  Here we present updated results using more sensitive
70$\mu$m data.

The sample of 427 xFLS-70 sources with redshifts and with only one
candidate 24$\mu$m counterpart (Sec.\,4.2) was matched to the
$4\sigma$ radio catalog (Condon et al. 2003).  Within $3\arcsec$, 325
of these xFLS-70 sources (76\%) have radio counterparts.  Figure 11
shows the observed S70/S(20\,cm) ratio as a function of redshift for
the matched galaxies.

At low redshift, the FIR to radio $q$-parameter is defined as
$q\equiv{\rm Log}(FIR/(3.75\times10^{12} {\rm W\,m}^{-2})) - {\rm
Log}(S(20{\rm cm})/({\rm W\,m}^{-2}{\rm Hz}^{-1}))=2.3\pm0.2$ (Helou
et al. 1985).  Based on the typical SED of the xFLS sample of galaxies
(Sec.\,4.2), the corresponding predicted $q$ parameter for the
70$\mu$m to radio flux density ratio is $q_{70}\equiv {\rm
Log}(S70/S[20{\rm cm}]) =2.09$ at $z\simeq 0.2$.  We measure an
average value of $q_{70}=2.10\pm0.16$ for xFLS galaxies at $z\simeq
0.2$.  By comparison, Appleton et al. (2004) derived a similar value
of $q_{70}=2.16\pm0.17$, but unlike the previous results we observe
the expected decrease of observed $q_{70}$ as a function of redshift
(Fig. 11).  Based on the SEDs of local galaxies, we expect the
observed $q_{70}$ parameter to decrease with redshift since the
average IR spectral index is significantly steeper than the radio
spectral index.  Assuming an IR spectral index of $\alpha=2.4$ based
on the observed average ratio of S70/S24$=14$ and the typical
non-thermal radio spectral index of $\alpha=0.8$ ($f_{\nu} \propto
\nu^{-\alpha}$, Condon et al. 1992), we expect $q_{70} \propto
(1+z)^{-1.6}$.  From a least-squares fit to the data, we estimate
$q_{70} \propto (1+z)^{-1.4\pm0.6}$.  Hence, the observed $q_{70}$ for
the xFLS galaxies follows the expected trend out to at least $z\sim 1$
(Fig. 11).

\section{Conclusions}

The {\it Spitzer} 70$\mu$m and 160$\mu$m observations of the xFLS are
presented.  For the deeper verification field data, we measure number
counts down to 8\,mJy and 50\,mJy ($5\sigma$) in the 70$\mu$m and
160$\mu$m bands, respectively.  The observed xFLS counts are
consistent with previous measurements (Dole et al. 2004a) and are
consistent within the uncertainties with the evolutionary models
(Lagache et al. 2004).  Based on the models of Lagache et al. (2004),
approximately 35\% of the CIB is resolved at 70$\mu$m and 15\% is
resolved at 160$\mu$m at the depth of the verification data.  The
observed fraction of low redshift galaxies at bright 70$\mu$m flux
densities is larger than model predictions and the total counts appear
slightly systematically lower.  These results may suggest the models
overestimate the contribution of high-redshift galaxies at bright
70$\mu$m flux densities to the total CIB.  Deeper 70$\mu$m
observations are needed for measuring the expected turn over in the
differential source counts to provide better constraints on the
evolutionary models.

The observed xFLS infrared colors S70/S24 and S160/S70 are consistent
with the results from the {\it IRAS} and {\it ISO} missions.  Modeled
SED fits suggest an average 160$\mu$m/70$\mu$m color temperature for
the dust of $T_{d}\simeq 30$\,K, for the 160$\mu$m-selected sample of
galaxies.  This temperature is consistent with {\it ISO} observations
of spirals (Bendo et al. 2003).  The average S70/S24 ratio implies an
infrared spectral index of $\alpha\simeq 2.4$, which agrees with
expectations from the average {\it IRAS} S60/S24 ratio (Soifer et
al. 1989; Sanders et al. 2003).  The observed 70$\mu$m infrared to
radio correlation of the xFLS sources also agrees well with the
FIR-to-radio correlation found for local star-forming galaxies (Helou
et al. 1985; Condon et al. 1992).  We observe a trend of decreasing
S70/S(20\,cm) as a function of redshift consistent with expectations
based on the SEDs of local galaxies.

Given the lack of sufficient redshift measurements for the
high-redshift faint xFLS-Ge sources, we can only derive an infrared
luminosity function at low redshift for only the brightest 70$\mu$m
galaxies.  With these data, we calculate a rest-frame 60$\mu$m
luminosity function for the xFLS that agrees well with the {\it IRAS}
luminosity function (Soifer et al. 1987).  In the future, it may be
possible to measure the evolution of the luminosity function out to
moderate redshifts with the xFLS data when additional redshift surveys
of faint 70$\mu$m sources become available.

We thank all of our colleagues associated with the {\it Spitzer}
mission who have made these observations possible.  This work is based
on observations made with the {\it Spitzer Space Telescope}, which is
operated by the Jet Propulsion Laboratory, California Institute of
Technology under NASA contract 1407.

\epsscale{0.70}

\begin{figure}
\plotone{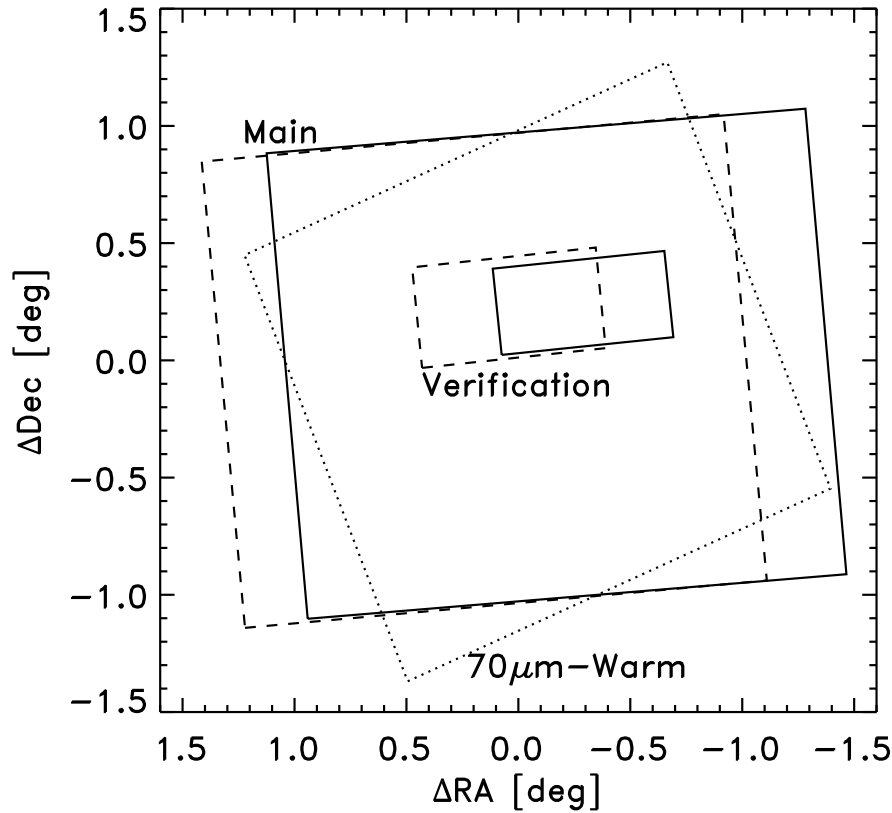}
\caption{The layout of the 70$\mu$m and 160$\mu$m fields for the xFLS
observations.  The 70$\mu$m main and verification fields are shown by
the solid-lines, while the 160$\mu$m xFLS fields are shown as
dashed-lines.  The location of the 70$\mu$m-warm observations taken in
2005 May is shown by the dotted-line. The right ascension and
declination offsets are with respect to the center of the main field
($17^{\rm h}18^{\rm m}00^{\rm s}$,
$+59^{\circ}30^{\prime}00^{\prime\prime}$).}
\end{figure}

\begin{figure}
\plotone{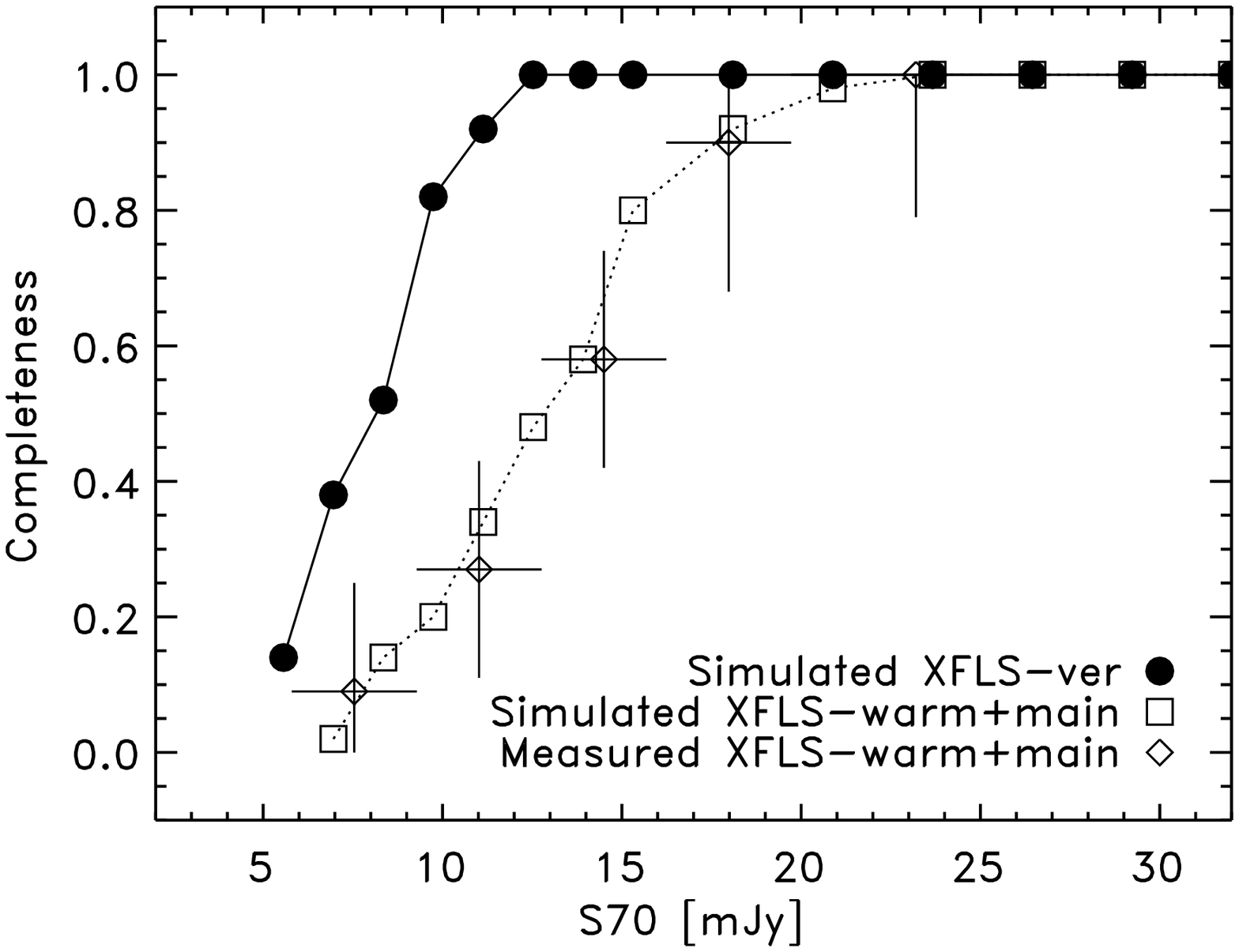}
\caption{The 70$\mu$m completeness level as a function of flux density
for the verification and warm+main xFLS surveys.  The diamonds
represent empirical measurements of the warm+main survey based on the
deeper verification data, and the squares and solid circles are the
results of simulations for the warm+main and verification fields
respectively.  The completeness estimates are for the S/N$>5$ source
lists used to derive the counts.}
\end{figure}

\begin{figure}
\plotone{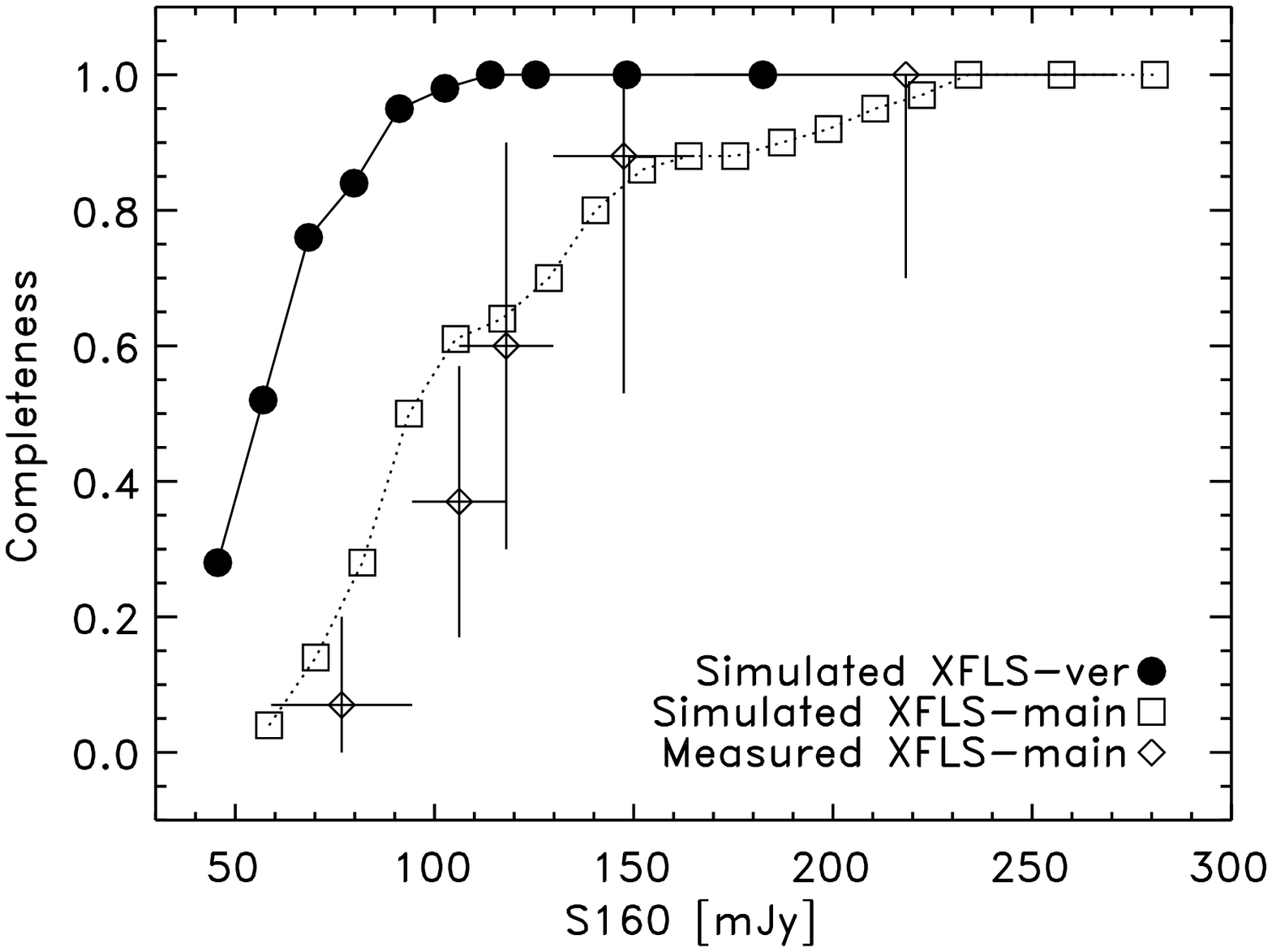}
\caption{The 160$\mu$m completeness level as a function of flux
density for the verification and main xFLS surveys.  The diamonds
represent empirical measurements of the main survey based on the
deeper verification data, and the squares and solid circles are the
results of simulations for the main and verification fields
respectively. The completeness estimates are for the S/N$>5$ source
lists used to derive the counts.}

\end{figure}

\begin{figure}
\plotone{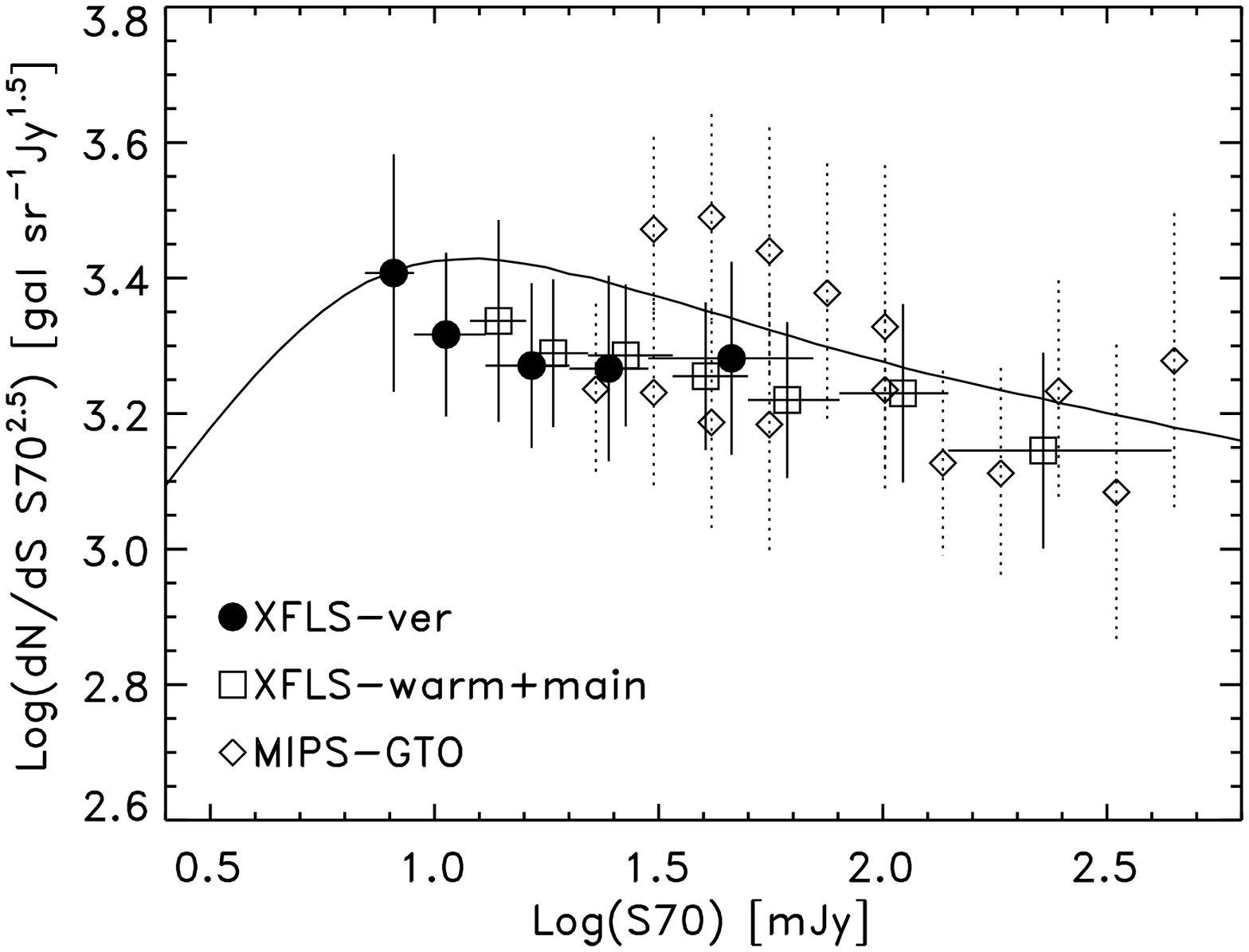}
\caption{The 70$\mu$m differential number counts for the verification
(solid circles) and warm+main (squares) xFLS surveys.  The diamonds
represent previous MIPS-GTO measurements (Dole et al. 2004a).  The
vertical error bars include the uncertainties of the absolute flux
density scale and completeness, and the horizontal line segments show
the bin sizes.  The solid line is the evolutionary model of Lagache et
al. (2004).}
\end{figure}

\begin{figure}
\plotone{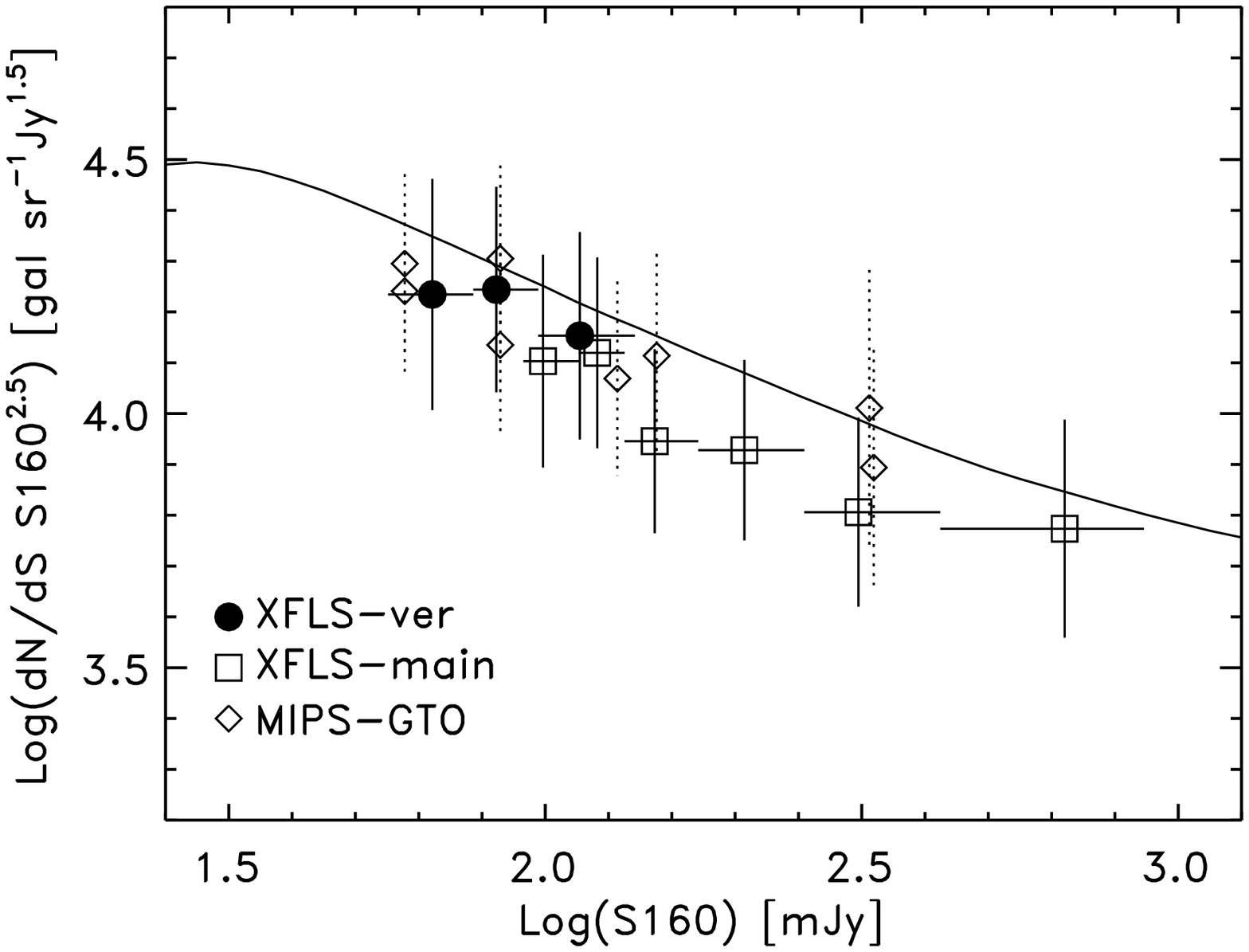}
\caption{The 160$\mu$m differential number counts for the verification
(solid circles) and main (squares) xFLS surveys.  The diamonds
represent previous MIPS-GTO measurements (Dole et al. 2004a).  The
vertical error bars include the uncertainties of the absolute flux
density calibration scale and completeness, and the horizontal line
segments show the bin sizes.  The solid line is the evolutionary model
of Lagache et al. (2004).}
\end{figure}

\begin{figure}
\plotone{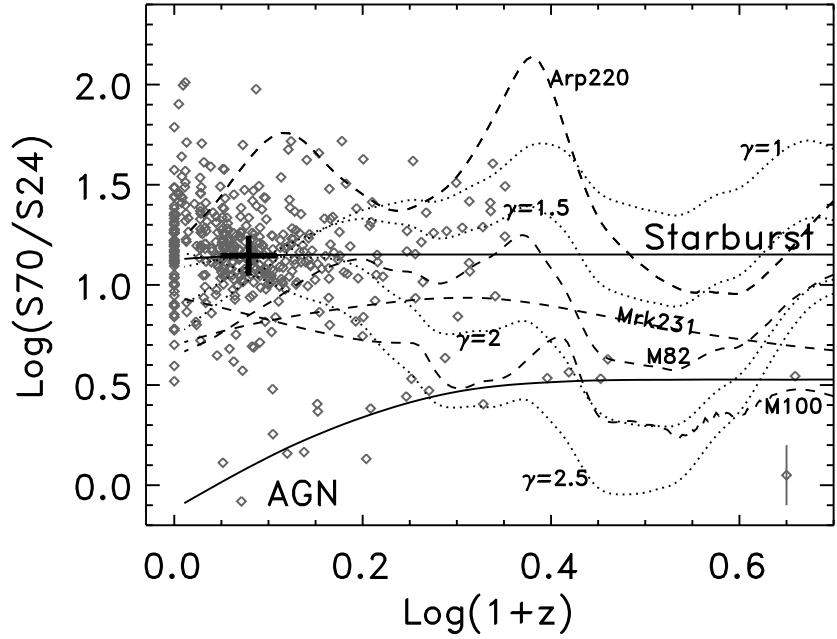}
\caption{The S70/S24 flux density ratio for 70$\mu$m sources with
redshifts.  The ULIRGs Arp\,220 and Mrk\,231, the starburst M82, and
the spiral galaxy M100 are shown as dashed lines.  The solid lines
represent simple modified blackbody SED models for starbursts
($\alpha=2.4$, $T_d=30$\,K) and AGN ($\alpha=1.1$, $T_d=90$\,K).  The
dotted lines show the SED models of Dale \& Helou (2002) as a function
of $\gamma$, where $\gamma\sim 2$--2.5 represent more quiescent
star-forming galaxies and $\gamma\sim 1.5$--1 represent more active
galaxies.  The plus symbol shows the predicted S70/S24 ratio at a
redshift $z=0.2$ based on the average {\it IRAS} S60/S25 ratio.  The
typical error bar for the data points from the $4\sigma$ source lists
is shown in the lower right.}
\end{figure}

\begin{figure}
\plotone{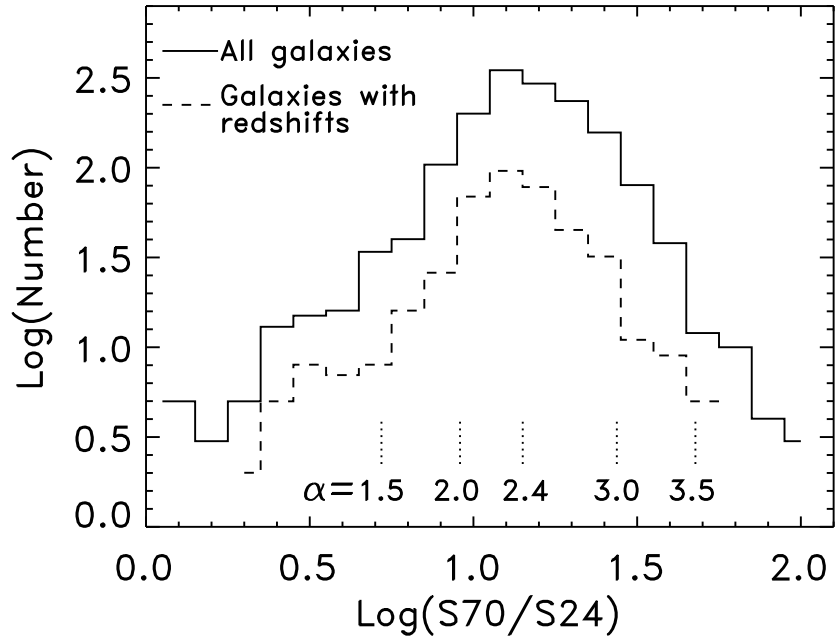}
\caption{The distributions of the S70/S24 flux density ratios for
70$\mu$m sources with redshifts shown in Fig. 6 (dashed line) and for
all 70$\mu$m-selected sources (solid line).  Values for a range of
infrared spectral indexes ($\alpha$) are shown by the vertical dotted
lines.}
\end{figure}

\begin{figure}
\plotone{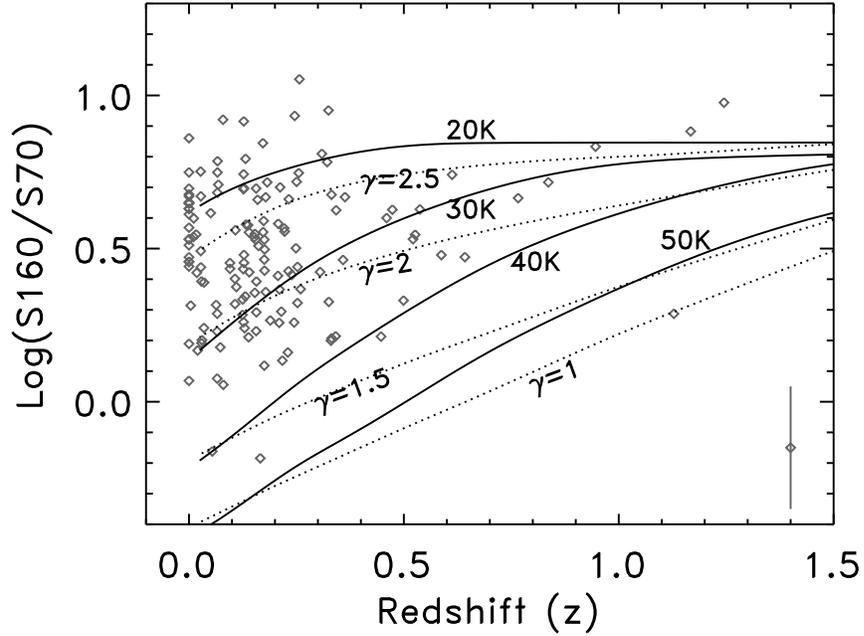}
\caption{The S160/S70 flux density ratio for 160$\mu$m-selected
sources with redshifts.  The solid lines show simple modified
blackbody SED models based on different dust temperatures (20--50\,K)
assuming an infrared spectral index of $\alpha =2.4$ and a dust
emissivity index of $\beta=1.5$.  The dotted lines show the SED models
of Dale \& Helou (2002) as a function of $\gamma$.  The typical error
bar for the data points from the $4\sigma$ source lists is shown in
the lower right.}
\end{figure}

\begin{figure}
\plotone{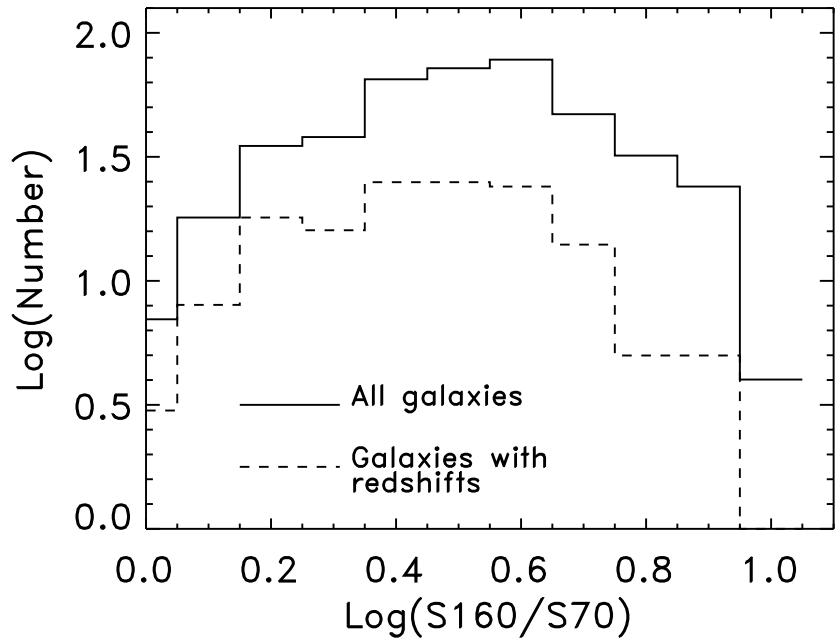}
\caption{The distributions of the S160/S70 flux density ratios for the
160$\mu$m sources with redshifts shown in Fig. 8 (dashed line) and for
all 160$\mu$m-selected sources (solid line).  The excess of sources
without redshifts at the higher S160/S70 ratios may suggest that the
current redshift sample of 160$\mu$m sources is biased toward
lower redshifts.}
\end{figure}

\begin{figure}
\plotone{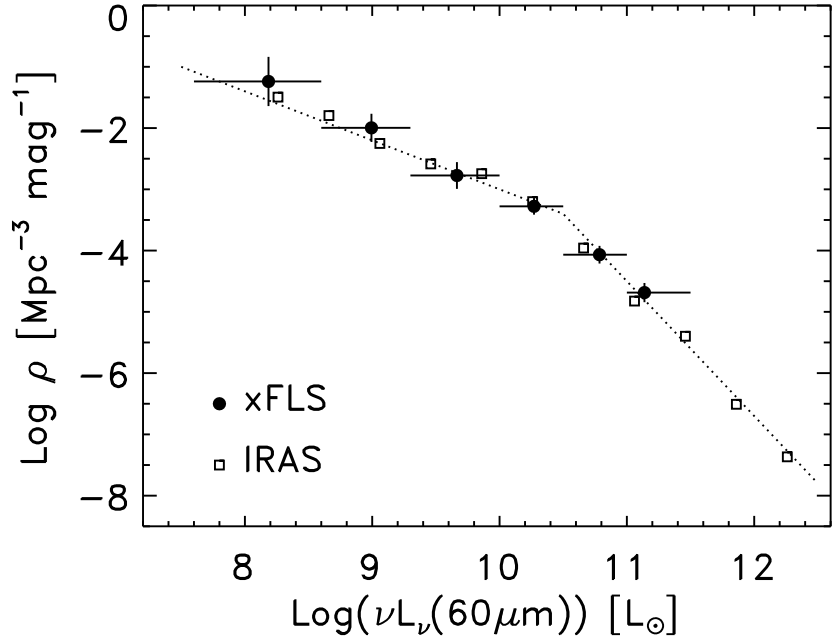}
\caption{The xFLS rest-frame 60$\mu$m luminosity function based on the
sources with S70$>50$\,mJy and $z<0.3$ (solid circles).  The {\it
IRAS} 60$\mu$m luminosity function is represented by the squares
(Soifer et al. 1987), corrected for the adopted value of
H$_0=70\kps\,{\rm Mpc}^{-1}$.  The dotted-line shows the double
power-law fit of the luminosity function based on the {\it IRAS} data
(Soifer et al. 1987; Sanders et al. 2003).}
\end{figure}

\begin{figure}
\plotone{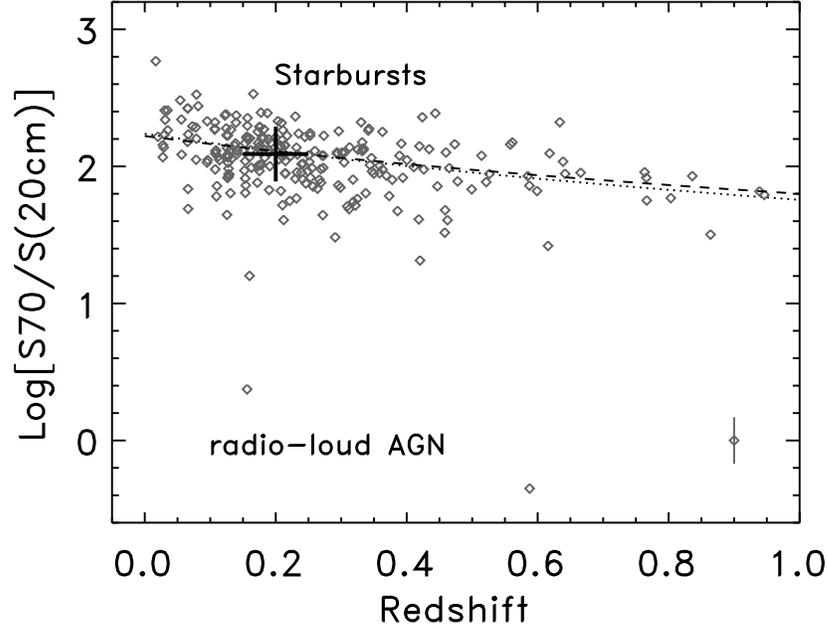}
\caption{The S70/S(20\,cm) flux density ratio for the
70$\mu$m-selected xFLS sources with redshifts between $0 < z < 1$.
The dotted-line shows the expected behavior of S70/S(20\,cm)$\propto
(1+z)^{-1.6}$, assuming power-law approximations for the SEDs at radio
and infrared wavelengths.  The dashed-line shows the result of a
least-squares fit [S70/S(20\,cm)$\propto (1+z)^{-1.4\pm0.6}$]. The plus
symbol shows the predicted value of the S70/S(20\,cm) ratio based on
the local infrared to radio correlation.  The typical error bar for
the data points from the $4\sigma$ source lists is shown in the lower
right.}

\end{figure}

\begin{deluxetable}{lcccccc}
\tablecaption{{\it Spitzer} Extragalactic FLS MIPS-Ge Observations}
\tablehead{
\colhead{} &
\multicolumn{2}{c}{$t_{int}$ (s)} &
\colhead{}&
\multicolumn{2}{c}{$5\sigma$ (mJy)} & Area\\
\cline{2-3} \cline{5-6} 
\colhead{Field} &
\colhead{70$\mu$m}&
\colhead{160$\mu$m}&
\colhead{} &
\colhead{70$\mu$m}&
\colhead{160$\mu$m} & (deg$^2$)}
\startdata
Main\dotfill         & 42  &  8 &  & 19 & 100 & 4.5 \\
Verification\dotfill & 252 & 42 &  &  8 & 50 & 0.35\\
Warm+Main\dotfill    & 84  &  \dotfill &  & 14 & \dotfill & 3.3 \\
\enddata

\tablecomments{The integration time ($t_{int}$) is the average
integration time for the field.  The approximate average $5\sigma$
noise levels are for point sources and do not include the absolute
flux density calibration uncertainties.}

\end{deluxetable}

\begin{deluxetable}{lccccc}
\tablecaption{Extragalactic FLS 70$\mu\lowercase{m}$ Catalog}
\tablehead{
\colhead{Source Name}&
\colhead{$\alpha$(J2000)} &
\colhead{$\delta$(J2000)} &
\colhead{Error} &
\colhead{S70}&
\colhead{Flag}\\
&\colhead{(deg)}&
\colhead{(deg)}&
\colhead{(arcsec)}&
\colhead{(mJy)}&
\colhead{ }\\
\colhead{(1)}&
\colhead{(2)}&
\colhead{(3)}&
\colhead{(4)}&
\colhead{(5)}&
\colhead{(6)}
}
\startdata

FLS70 J172723.3+601626& 261.84712& 60.27414& 1.9&   81.4$\pm$12.9 & prf\\
FLS70 J172716.2+600051& 261.81791& 60.01433& 1.6& 1610.1$\pm$244.6 & a120m\\
FLS70 J172709.5+600141& 261.78960& 60.02815& 2.1&   77.3$\pm$12.7 & prf\\
FLS70 J172704.7+593736& 261.76978& 59.62680& 2.0&   62.1$\pm$9.9 & prf\\
FLS70 J172651.1+601851& 261.71329& 60.31419& 3.3&   34.7$\pm$7.0 & prf\\

\enddata

\tablecomments{The entire xFLS 70$\mu$m catalog ($>7\sigma$) is
presented in the electronic edition of the Astronomical Journal. A
portion of Table 2 is shown here for guidance regarding the form and
content of the catalog.  Column (1) gives the source name following
the IAU designations (Lortet, Borde, \& Ochsenbein 1994).  Column (2)
and column (3) are the right ascension ($\alpha$) and declination
($\delta$) J2000.0 source positions in decimal degrees.  Column (4)
provides the radial positional uncertainty in arcsec ($1\sigma$).
Column (5) gives the flux density measurement in mJy.  The flux
density errors include the uncertainties of the absolute flux density
scale.  Column (6) provides a flag for the flux density measurement
method.  A flag of ``prf'' indicates that the source was fitted by the
PRF, while a flag starting with an ``a'' indicates that an aperture
measurement was used (for cases of extended sources and/or close
blends).  The number following the ``a'' gives the aperture diameter
in arcsec; e.g., a120 implies an aperture diameter of $120\arcsec$.
Sources with aperture flags ending with a ``m'' represent bright
galaxies not well fitted by the PRF and with nearby companions; in
these cases, the flux density of the bright galaxy was derived by
subtracting the PRF measurement(s) of the nearby faint point source(s)
from the total aperture flux density.}

\end{deluxetable}

\begin{deluxetable}{lccccc}
\tablecaption{Extragalactic FLS 160$\mu\lowercase{m}$ Catalog}
\tablehead{
\colhead{Source Name}&
\colhead{$\alpha$(J2000)} &
\colhead{$\delta$(J2000)} &
\colhead{Error} &
\colhead{S160}&
\colhead{Flag}\\
&\colhead{(deg)}&
\colhead{(deg)}&
\colhead{(arcsec)}&
\colhead{(mJy)}&
\colhead{ }\\
\colhead{(1)}&
\colhead{(2)}&
\colhead{(3)}&
\colhead{(4)}&
\colhead{(5)}&
\colhead{(6)}
}
\startdata

FLS160 J172902.1+600523& 262.25801& 60.08859& 3.4&  294.8$\pm$76.0& prf\\
FLS160 J172855.5+600026& 262.23039& 60.00619& 7.4&  154.4$\pm$44.2& prf\\
FLS160 J172851.3+600821& 262.21295& 60.13807& 3.8&  293.7$\pm$76.4& prf\\
FLS160 J172826.5+600539& 262.10963& 60.09320& 3.0& 1211.0$\pm$310.5& a192\\
FLS160 J172817.7+600100& 262.07271& 60.01573& 5.9&  188.4$\pm$50.97& prf\\

\enddata

\tablecomments{The entire xFLS 160$\mu$m catalog ($>7\sigma$) is
presented in the electronic edition of the Astronomical Journal. A
portion of Table 3 is shown here for guidance regarding the form and
content of the catalog.  Column headers are the same as described in
Table 2. }

\end{deluxetable}

\begin{deluxetable}{cccccc}
\tablecaption{Extragalactic FLS 70$\mu\lowercase{m}$ Number Counts}
\tablehead{
\colhead{Average S$_{\nu}$} &
\colhead{S$_{\rm low}$}&
\colhead{S$_{\rm high}$} &
\colhead{Number} &
\colhead{Completeness} &
\colhead{$dN/dS\,S^{2.5}$}\\
\colhead{(mJy)}&
\colhead{(mJy)}&
\colhead{(mJy)}&
\colhead{}&
\colhead{}&
\colhead{(gal sr$^{-1}$ Jy$^{1.5}$)}
}
\startdata

  8.1&  7&  9& 46&  0.50$\pm$0.15 & 2556$\pm$1030\\
 10.6&  9& 13& 61&  0.80$\pm$0.08 & 2073$\pm$ 575\\
 16.5& 13& 20& 40&  1.0$\pm$0.05  & 1866$\pm$ 522\\
 24.5& 20& 30& 21&  1.0$\pm$0.03  & 1847$\pm$ 582\\
 46.0& 30& 70& 18&  1.0$\pm$0.03  & 1912$\pm$ 626\\
\ \\
 13.9& 12& 16& 230&  0.60$\pm$0.15 & 2172$\pm$744\\
 18.4& 16& 22& 230&  0.90$\pm$0.08 & 1883$\pm$453\\
 26.7& 22& 34& 200&  1.00$\pm$0.05 & 1931$\pm$466\\
 40.3& 34& 50&  89&  1.00$\pm$0.03 & 1800$\pm$451\\
 61.2& 50& 80&  54&  1.00$\pm$0.03 & 1660$\pm$439\\
110.9& 80& 140& 25&  1.00$\pm$0.03 & 1698$\pm$514\\
227.9&140& 440& 17&  1.00$\pm$0.03 & 1398$\pm$464\\

\enddata

\tablecomments{The first five entries are the 70$\mu$m counts for the
verification field, while the remaining entries are for the warm+main
70$\mu$m survey.  The uncertainties for the differential counts
($dN/dS\,S^{2.5}$) include the Poisson noise, the completeness
uncertainty, and the absolute flux density calibration uncertainty of
15\%.  The counts are based on the $>5\sigma$ source lists, corrected
for completeness.}

\end{deluxetable}

\begin{deluxetable}{cccccc}
\tablecaption{Extragalactic FLS 160$\mu\lowercase{m}$ Number Counts}
\tablehead{
\colhead{Average S$_{\nu}$} &
\colhead{S$_{\rm low}$}&
\colhead{S$_{\rm high}$} &
\colhead{Number} &
\colhead{Completeness} &
\colhead{$dN/dS\,S^{2.5}$}\\
\colhead{(mJy)}&
\colhead{(mJy)}&
\colhead{(mJy)}&
\colhead{}&
\colhead{}&
\colhead{(gal sr$^{-1}$ Jy$^{1.5}$)}
}
\startdata

68.0&  56.4& 77.0&  17&   0.55$\pm$0.15& 17156$\pm$ 8978\\
85.9&  77.0& 97.5&  15&   0.85$\pm$0.08& 17541$\pm$ 8155\\
116.4& 97.5& 138.6&  13&   0.97$\pm$0.05& 14231$\pm$ 6678\\
\ \\
101.8&  92.4& 112.9 &  60&   0.55$\pm$0.15 & 12682$\pm$ 6104\\
124.0& 112.9& 133.4 &  45&   0.65$\pm$0.1 & 13170$\pm$ 5688\\
152.8& 133.4& 174.5 &  44&   0.80$\pm$0.08& 8825$\pm$ 3674\\
211.7& 174.5& 256.6 &  43&   0.92$\pm$0.05& 8472$\pm$ 3460\\
320.7& 256.6& 420.8 &  25&   1.0$\pm$0.05 & 6399$\pm$ 2738\\
679.0& 420.8& 882.6 &  10&   1.0$\pm$0.05 & 5935$\pm$ 2926\\

\enddata

\tablecomments{The first three entries are the 160$\mu$m counts for
the verification field, while the remaining entries are for the main
160$\mu$m survey.  The uncertainties for the differential counts
($dN/dS\,S^{2.5}$) include the Poisson noise, the completeness
uncertainty, and the absolute flux density calibration uncertainty of
25\%.  The counts are based on the $>5\sigma$ source lists, corrected
for completeness.}

\end{deluxetable}

\end{document}